\documentclass[final,5p,times,twocolumn,number]{elsarticle}
\usepackage[utf8]{inputenc}
\usepackage{array,graphicx,amsfonts,amssymb,amsmath,mathrsfs,hyperref,bm,dsfont,mathtools}
\usepackage{stackengine}
\usepackage{relsize}
\usepackage{comment}
\usepackage{physics}
\usepackage{braket}
\usepackage{ulem}
\usepackage[export]{adjustbox}
\usepackage{todonotes}

\begin{document}

\begin{frontmatter}

\title{Order-by-disorder and long-range interactions in the antiferromagnetic transverse-field Ising model on the triangular lattice - A perturbative point of view}
\author[a]{Jan Alexander Koziol}
\author[a]{Matthias Mühlhauser}
\author[a]{Kai Phillip Schmidt}
\affiliation[a]{organization={Friedrich-Alexander-Universität Erlangen-Nürnberg (FAU), Department Physik}, addressline={Staudtstraße 7}, postcode={D-91058 Erlangen}, state={Bavaria}, country={Germany}}

\begin{abstract}
		We study the low-field ground-state (GS) properties of the antiferromagnetic transverse-field Ising model with long-range interactions (afLRTFIM) on the triangular lattice.
We use the method of perturbative continuous unitary transformations (pCUT) to derive an effective model for the degenerate GS space of the antiferromagnetic nearest-neighbour (NN) Ising model on a finite system, by treating the transverse-field (TF) and the long-range interactions (LRI) as a perturbation.
We determine a level-crossing between the plain stripe phase at small TF and the clock-ordered phase at intermediate TF at $h\cong0.129$ for $\alpha=6$, $N=36$ spins in order three perturbation theory. 
We discuss the qualitative layout of the quantum phase diagram of the afLRTFIM on the triangular lattice.
\end{abstract}


\end{frontmatter}

{\textit{Introduction.-}}    
Geometric frustration in lattice models is a common ingredient to potentially trigger emergent exotic quantum behaviour \cite{Moessner2000,Moessner2001,Moessner2006}.
Similarly, considering LRI also gives potential rise to new GS properties compared to a short-range interacting model \cite{Scholl2021,Koziol2023}. 
This work aims to report an observation on the interplay between LRI and mechanisms arising from geometric frustration.

{\textit{Model.-}}    
The workhorse model for our demonstration is the afLRTFIM on the triangular lattice 
\begin{align}
		\label{eq:lrtfim}
		H = \sum_{\langle i,j\rangle}\sigma_i^z\sigma_j^z+\frac{\lambda}{2}\sum_{i\neq j}\frac{(1-\chi_{\langle i,j \rangle})}{|\vec r_i -\vec r_j|^\alpha}\sigma_i^z\sigma_j^z+h\sum_i \sigma_i^x
\end{align}
with Pauli matrices $\sigma_i^{z/x}$ describing spins $1/2$ on lattice sites $\vec r_i$, coupling $\lambda > 0$, $\chi_{\langle i,j\rangle}$ is $1$ if $i$ and $j$ are NNs and 0 else, and the TF $h>0$. 
The parameter $\alpha>2$ determines the algebraic decay of the LRI. 
The limit $\alpha=\infty$ recovers the NN transverse-field Ising model (TFIM). 
As the triangular lattice is non-bipartite, it is not possible to align the $\sigma_i^z$ spin directions on every link of the lattice antiferromagnetically. 
This results in the following rule: Every state that does not have any closed loop of length three with all spins having the same orientation is a GS of the model at $h=0$ and $\alpha=\infty$. 
This manifests itself in an extensive residual entropy $S/N=0.323066$ \cite{Wannier1950,Wannier1973}.
It was demonstrated for $h>0$ and $\alpha=\infty$ that an order-by-disorder scenario breaks the GS degeneracy for $h>0$ and an emergent gapped \mbox{$\sqrt{3}\times\hspace{-3pt}\sqrt{3}$-clock} order is the GS \cite{Moessner2000,Moessner2001}.
With an increasing TF the clock order breaks down by a 3D-XY quantum phase transition into a trivial field-polarised phase at $h_c/J=1.65\pm0.05$ \cite{Isakov2003,Powalski2013}.
Regarding the effect of antiferromagnetic LRI on the degenerate GS space of the antiferromagnetic NN Ising model with $h=0$, it has been demonstrated recently that a six-fold degenerate gapped plain stripe pattern is the GS for $\alpha<\infty$ \cite{Koziol2019,Koziol2023}. %

{\textit{Method.-}}    
We set up a degenerate perturbation theory calculation using pCUT \cite{Knetter2000,Knetter2003} treating the TF and the LRI as a perturbation on the degenerate NN Ising GS space. 
As the number of GSs grows exponentially with the system size \cite{Wannier1950,Wannier1973}, it is not feasible to perform a calculation in the thermodynamic limit. 
Here, we consider a finite system of $N=6\times 6=36$ spins. 
In order to display the expected GSs, the linear system size needs to be a multiple of six \cite{Moessner2000,Moessner2001,Koziol2023}.
The next larger possible system cannot be handled by the numerical procedure.
To better approximate the thermodynamic limit on the finite system, we use resummed couplings \cite{Koziol2023}
\begin{align}
		\tilde J^{K,\alpha}_{i,j} = \sum_{k=-K}^{K} \sum_{l=-K}^{K} \frac{1}{|\vec r_i - \vec r_j+l \vec T_1 - k \vec T_2|^\alpha}
\end{align}
with a cutoff $K$ and the translational vectors of the unit cell $\vec T_1=(6,0)^T$ and $\vec T_2=(3,3\sqrt{3})^T$. 
We extrapolate the $\tilde J^{K,\alpha}_{i,j}$ in the finite cutoff $K$ to infinity $\tilde J^{\infty,\alpha}_{i,j}$ \cite{Koziol2023}. 
For $\alpha=\infty$, this is equivalent to periodic boundary conditions.
We encode the GSs in an augmented representation including the spin degrees of freedom (DOF) to deal with the LRI and link DOF defined by the two orientations of NN spins $\sigma_i^z\sigma_j^z$ to efficiently treat the TF
\begin{equation}
\label{eq:state}
\bigotimes_{i=0}^{N-1}\ket{(\sigma_i^z,\sigma_i^z\sigma_{\delta_1(i)}^z),(\sigma_i^z,\sigma_i^z\sigma_{\delta_2(i)}^z),(\sigma_i^z,\sigma_i^z\sigma_{\delta_3(i)}^z)}
\end{equation}
with $\delta_{1/2/3}(i)$ denoting three unique NN of $i$ such that each bond of the lattice is included exactly once in the state. 
In the augmented representation, Eq.~\eqref{eq:lrtfim} reads for the finite system
\begin{align}
		\nonumber H = & \sum_{i=0}^{N-1}\left[\sum_{\nu=0}^2(1,\tau^z)_{i+\nu}+h\prod_{\nu=0}^2(\sigma^x,\tau^x)_{i+\nu}\prod_{\kappa\in\Gamma_i}(1,\tau^x)_{\kappa}\right] \\
			& + \lambda \sum_{i=0}^{N-1}\sum_{j=0}^{N-1} \left[\frac{\tilde J^{\infty,\alpha}_{i,j}}{2}-\frac{\chi_{\langle i,j\rangle}}{2}\right] (\sigma^z,1)_i(\sigma^z,1)_j
\end{align}
with Pauli matrices $\tau^{x/z}$ acting on the link DOF and $\Gamma_i$ the set of the three indices to the NN bonds of $i$ which are not addressed by the $\delta_{1/2/3}(i)$ function.
To perform the perturbation theory around the NN Ising limit, we consider $\sum_{i=0}^{N-1}\sum_{\nu=0}^2(1,\tau^z)_{i+\nu}$ as the unperturbed Hamiltonian $H_0$.
The GSs of $H_0$ have the energy $E_0=-36$ and every link deviating from the two-one rule of the considered GS subspace provides an energy of $+4$.
We can represent the perturbation in terms of $T_n$-operators with $n\in \mathbb{Z}$. 
A $T_n$ operator changes the unperturbed state into a state with an energy difference (quasi-particle number) of $n$.
Note, the quasi-particles are associated to the link DOF (see structure of $H_0$).
The LRI becomes a $T_0$ operator and the TF is decomposed in $T_0$, $T_{\pm 4}$, $T_{\pm 8}$, and $T_{\pm 12}$ operators.
The structure of $H_0$ and the perturbation makes the pCUT approach applicable \cite{Knetter2000, Knetter2003}.
The method perturbatively determines a block-diagonal effective Hamiltonian for each quasi-particle number including the GS block. 
We evaluate the matrix elements for the effective GS model (zero quasi-particle block) up to order three in $h$ and $\lambda$. 
Further, we insert values for the perturbation parameters and diagonalise the matrix of the effective model.

\begin{figure}
	\centering
	\includegraphics[]{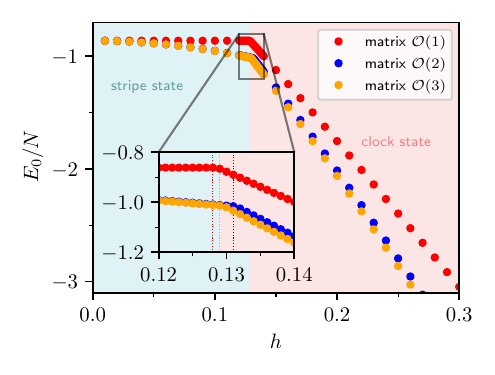}
	\caption{GS energy per site for $\alpha=6$. Background colors indicate the nature of the GS. Inset: Zoom into the region of the level crossing. Dashed vertical lines indicate the transition between the two GSs.}
	\label{fig}
\end{figure}

{\textit{Results.-}}
For $h=0$, there is only a contribution in first order in $\lambda$ since the perturbation consists in this case only of a $T_0$ operator.
For $\lambda=0$, the effective model in first order couples $h=0$ states, where spins can be flipped without an energy cost \cite{Moessner2000,Moessner2001}.
Note, the TF favours states the most with a maximal number of these flippable spins.
On the other hand, the $h=0$ stripe states for $\lambda=1$ and $\alpha<\infty$ do not contain any of these local configurations.
Therefore the LRI and the TF beneficiate different subsets of states from the unperturbed degenerate GS space.
In the following, we focus on $\alpha=6$ motivated by experimental realisations using laser-driven Rydberg atoms \cite{Scholl2021}.
We set $\lambda=1$ as the largest contribution to the perturbation due to the LRI is of the order of $0.04$ for $\alpha=6$.
We evaluate the GS energy $E_0$ numerically for given parameters $h$, by calculating the smallest eigenvalue of the derived effective model.
We present the determined GS energy per site for $\alpha=6$ in Fig.~\ref{fig}. 
We report a level crossing at $h\cong 0.129$ between the plain stripe state at small $h$ and a clock-ordered state at intermediate $h$.
We observe a convergence for the second and third order energy values compared to the first order.
The transition values for the different orders change non-monotonously and convergence of the energy for the stripe state is better than in the fluctuation driven clock-ordered state.

{\textit{Conclusions.-}}
From literature, it is known that the quantum phase diagram of the afLRTFIM for $\alpha<\infty$ consists of a sixfold degenerate low-TF plain stripe phase \cite{Koziol2023}, an intermediate clock-ordered phase, and a high-TF $x$-polarised phase \cite{Humeniuk2016,Fey2019,Koziol2019}.
It was demonstrated in Ref.~\cite{Fey2019} that the phase transition between the $x$-polarised phase and the clock-ordered phase remains of 3D-XY universality for $\alpha<\infty$ and the $h_c$ value decreases with $\alpha$ \cite{Humeniuk2016,Fey2019}.
With the method discussed above it is possible to calculate estimators for the $h_c$ between the plain stripe low-field phase and the clock-ordered phase using a single perturbative ansatz. 
Similar calculations were already performed for triangular lattice cylinder geometries \cite{Koziol2019}, but there a separate perturbative calculation was performed for the stripe and the clock-ordered phase.
It remains an open research question if the intermediate clock-ordered phase persists for small values $\alpha\lesssim 3$ or if there is a direct transition to the stripe phase \cite{Saadatmand2018,Fey2019,Koziol2019}.
This question cannot be answered with the method presented above, since it stretches beyond the convergence of the perturbative argument.
Nevertheless, the approach is suitable for investigations at large $\alpha$ values, e.\,g. $\alpha=6$, which makes it a suitable tool to gauge low-TF transitions for experimental realisations with Rydberg atoms \cite{Scholl2021} or other quantum simulators.

{\textit{Funding.-}}
This work was supported by the Deutsche Forschungsgemeinschaft (DFG, German Research Foundation) -- Project-ID 429529648—TRR 306 \mbox{QuCoLiMa} (``Quantum Cooperativity of Light and Matter'').\\

{\textit{CRediT author contribution statement.-}}
J. A. Koziol: Conceptualization, Data curation, Formal analysis, Software, Investigation, Methodology, Writing - original draft. M. Mühlhauser: Methodology, Software, Writing - review \& editing. K. P. Schmidt: Supervision, Methodology, Writing - review \& editing

{\textit{Declaration of competing interest.-}}
The authors declare that they have no known competing financial interests or personal relationships that could have appeared to influence the work reported in this paper.


%

\end{document}